\documentclass[sigconf]{acmart}

\usepackage{subcaption}
\usepackage{xcolor}
\usepackage{soul} 
\soulregister\cite7
\soulregister\ref7

\AtBeginDocument{%
  \providecommand\BibTeX{{%
    \normalfont B\kern-0.5em{\scshape i\kern-0.25em b}\kern-0.8em\TeX}}}

\copyrightyear{2023} 
\acmYear{2023} 
\setcopyright{rightsretained}

\acmConference[Conference acronym 'XX]{Make sure to enter the correct
  conference title from your rights confirmation emai}{June 03--05,
  2018}{Woodstock, NY}
\acmBooktitle{Woodstock '18: ACM Symposium on Neural Gaze Detection,
 June 03--05, 2018, Woodstock, NY} 
\acmPrice{15.00}
\acmISBN{978-1-4503-XXXX-X/18/06}

\begin{document}

\title [Purrfect Pitch: Exploring Musical Interval Learning through Multisensory Interfaces]{Purrfect Pitch: Exploring Musical Interval Learning \newline through Multisensory Interfaces}

\settopmatter{authorsperrow=4} 
\author{Sam Chin}
\authornote{The authors contributed equally to this research.}
\email{chins@media.mit.edu}
\affiliation{
  \institution{MIT Media Lab}
  \city{Cambridge}
  \state{MA}
  \country{USA}
}

\author{Cathy Mengying Fang}
\authornotemark[1]
\email{catfang@media.mit.edu}
\affiliation{
  \institution{MIT Media Lab}
  \city{Cambridge}
  \state{MA}
  \country{USA}
}

\author{Nikhil Singh}
\authornotemark[1]
\email{nsingh1@mit.edu}
\affiliation{
  \institution{MIT Media Lab}
  \city{Cambridge}
  \state{MA}
  \country{USA}
}

\author{Ibrahim Ibrahim}
\email{ibrahim@gsd.harvard.edu}
\affiliation{
  \institution{Harvard Graduate School of Design}
  \city{Cambridge}
  \state{MA}
  \country{USA}
}

\author{Joe Paradiso}
\email{joep@media.mit.edu}
\affiliation{
  \institution{MIT Media Lab}
  \city{Cambridge}
  \state{MA}
  \country{USA}
}

\author{Pattie Maes}
\email{pattie@media.mit.edu}
\affiliation{
  \institution{MIT Media Lab}
  \city{Cambridge}
  \state{MA}
  \country{USA}
}

\renewcommand{\shortauthors}{Chin, Fang, and Singh et al.}


\begin{abstract}

We introduce \textit{Purrfect Pitch}, a system consisting of a wearable haptic device and a custom-designed learning interface for musical ear training. We focus on the ability to identify musical intervals (sequences of two musical notes), which is a perceptually ambiguous task that usually requires strenuous rote training. With our system, the user would hear a sequence of two tones while simultaneously receiving two corresponding vibrotactile stimuli on the back. Providing haptic feedback along the back makes the auditory distance between the two tones more salient, and the back-worn design is comfortable and unobtrusive. During training, the user receives multi-sensory feedback from our system and inputs their guessed interval value on our web-based learning interface. They see a green (otherwise red) screen for a correct guess with the correct interval value. Our study with 18 participants shows that our system enables novice learners to identify intervals more accurately and consistently than those who only received audio feedback, even after the haptic feedback is removed. We also share further insights on how to design a multisensory learning system.

% Multisensory approaches have shown promise in augmenting perceptual learning. In this work, we focus on the task of identifying musical intervals, which is a perceptually ambiguous task that usually requires strenuous rote training. We constructed a custom wearable haptic device accompanying a web-based learning interface. The user would listen to a musical interval (a sequence of two tones) while simultaneously receiving a vibrotactile stimulus on the back. In our study with 20 participants, we show that our system enables novice learners to identify intervals more accurately and consistently than those who only received audio, even after the haptic feedback is removed. We also share further insights on how to design a multisensory learning system.

% We demonstrate a multisensory approach for perceptual learning that supplements a perceptually-ambiguous audio stimulus with haptic feedback. In our study, participants hear a musical interval (a sequence of two tones) while simultaneously receiving a vibrotactile stimulus from a custom-designed wearable device. The results suggest that novice learners who experienced the multisensory auditory and haptic feedback could identify intervals more accurately and consistently than those who only received audio. Future work can leverage this insight to investigate the longer-term effects of multisensory learning, such as the retention of the ability to identify musical intervals after haptic feedback is removed.
  
\end{abstract}



\begin{teaserfigure}
\centering
  \includegraphics[width=.8\textwidth]{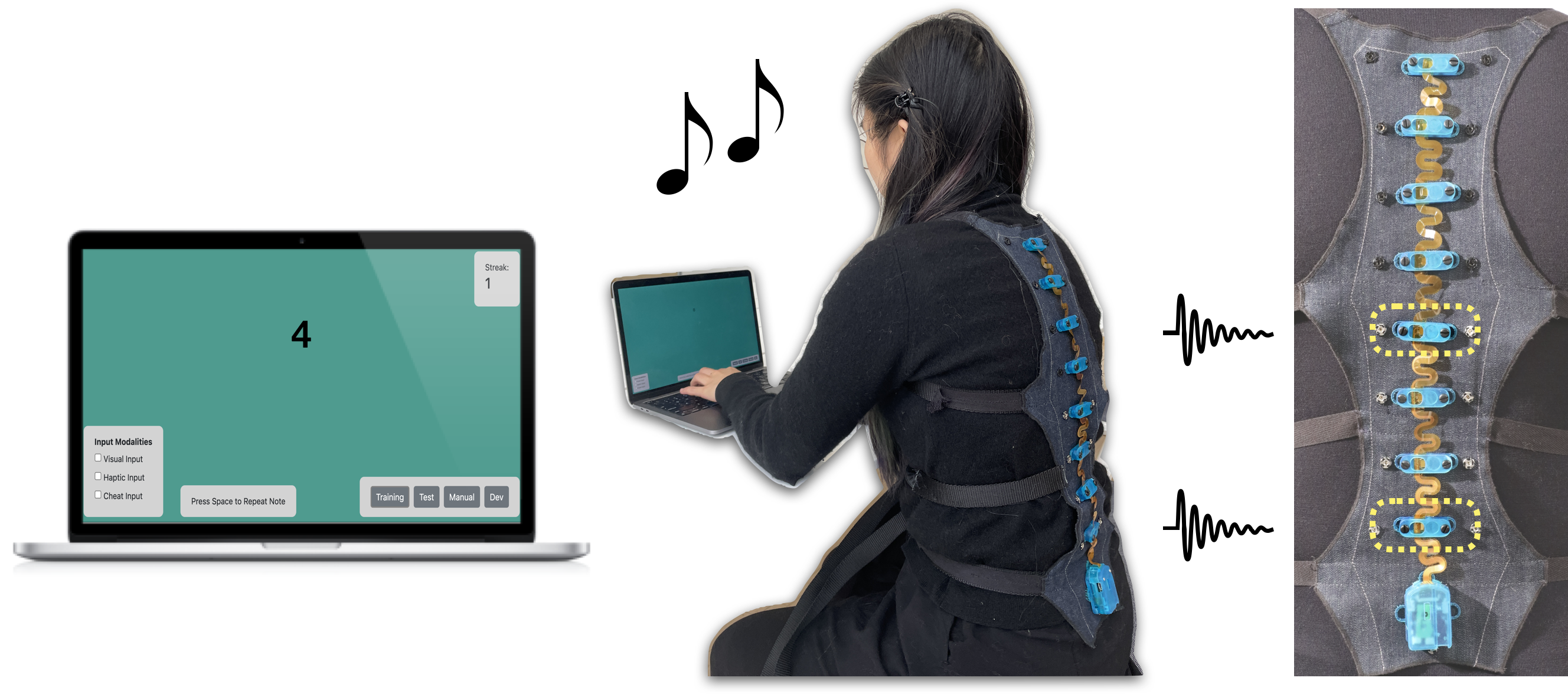}
  \caption{We developed a multisensory system that allows users to learn musical intervals. The system consists of a haptic device worn on the back (right) and a web-based learning interface (left). During learning, the user, for example, hears two musical notes, C and F, while feeling vibrations from two places on the back (highlighted in yellow). If they guessed the correct interval of 4, the interface displays a green screen (red if incorrect) and the correct numerical response.}
  \Description{xxxx}
  \label{fig:teaser}
\end{teaserfigure}


\maketitle

\section{Introduction}

Making sense of sensory information is the foundation of everyday activities. Profoundly, individuals develop improved acuity for sensory information through experience or practice in specific sensory tasks, which for some results in permanent changes in their perceptual response \cite{sep-perceptual-learning}. For example, some paint makers develop acuity to color and can mix paint from fundamental colors by simply looking; wine connoisseurs can differentiate notes in the smell and tastes of different wine types; professional piano tuners have more pitch acuity to fix out-of-tuned pianos.

% songwriters develop sensitivity to rhythms and notes in mundane life to produce unique songs. 

% Perceptual learning is defined as permanent learning that changes the perception as a result of practice or experience \cite{sep-perceptual-learning}. 

Different from learning math or physics, which is represented using symbols and abstractions, perceptual learning requires low-level changes and adaptation in how we respond to sensory stimuli. This is similar to learning a motor skill (e.g., learning how to bike or play sports), where completing bodily gestures is difficult to capture symbolically and thus hard to learn through written representations or observations alone \cite{kirsh2009knowledge, fang2024collography}. Both motor learning and perceptual learning often require strenuous practice to achieve mastery. 

To that end, a few systems have been developed to aid perceptual learning. While motor skill learning requires specific knowledge, perceptual learning can happen implicitly \cite{shmuelof2014recent}. Thus, multisensory perceptual learning offers an advantage by integrating information from multiple sensory modalities. To illustrate with an example, consider learning a new skill like ceramics. Learning to see when the clay is perfectly centered is primarily a visual task. However, learning to center the clay is much easier when you hear the rhythm of uneven spots and feel the clay pushing back against your hands. 
% Such as juggling, might be easier to learn if one can see, hear, and feel the motion of the objects they are juggling, rather than just using one modality
Similarly, when learning a new language, being able to hear the sounds of the words, see how they are written, and feel the movements of the mouth and tongue as they speak can all help learn the language more effectively.

% Multisensory perceptual learning refers to the process of learning a perceptual skill by integrating information from multiple sensory modalities, such as vision, hearing, and touch. 
% To illustrate with an example, consider learning a new skill like ceramics. Learning to see when the clay is perfectly centered is primarily a visual task. However, learning to  center the clay is much easier when you hear the rhythm of uneven spots and feel the clay pushing back against your hands. 
% Similarly, when learning a new language, being able to hear the sounds of the words, see how they are written, and feel the movements of the mouth and tongue as they speak can all help the individual learn the language more effectively

% We focus on learning auditory perceptual skills, in particular musical ear training. 

In our work, we focus on the specific case of learning auditory perceptual skills. In particular, we look at musical ear training, a common yet challenging task that is typically undertaken through only one modality. Ear training refers to gaining the ability to identify musical components (e.g., melodies and chords) by ear, an essential component of developing musical proficiency. For example, violinists need to tell the pitch based on hearing when needing to imitate recordings. Jazz musicians intuitively identify different chords and intervals to be able to improvise during a performance. However, identifying intervals and other musical elements by ear can be challenging for novices. Developing this skill requires extensive training, and the traditional rote approach of repetitive listening and identifying often requires long periods of strenuous, repetitive practice.

%CATHY: add some related work in assistive devices for musical ear training. others focused on rhythm, or what have you. 

In this paper, we demonstrate a multisensory (audio and haptic) platform for training to identify musical intervals, or the differences in pitch between two tones (two audio frequencies). We develop a perceptual training platform that involves an interface with a haptic wearable placed on the back. This device provides vibrotactile feedback concurrently while the notes are played. 
% [TODO EXPAND]
Our main research questions are: (1) can haptic reinforcement improve the performance of interval \textbf{recognition}? and (2) what is the effect on auditory perceptual \textbf{learning}?

% Our main research questions are: (1) Can multisensory learning using auditory and haptic modalities help individuals learn musical intervals more easily? (2) What are the relevant design considerations for developing a multisensory platform for learning the perception of musical intervals?

To understand whether haptic feedback has a positive effect on learning to perceive musical intervals, we conducted a study with 18 participants. Participants had no prior musical ear training, and about half were assigned to an audio-only control condition. Initial results show that participants with haptic feedback could identify musical intervals more accurately while feeling less frustrated and more engaged. 

% xxx
% xxx

In summary, this work's contribution is as follows:
\begin{itemize}
    \item The design of a multisensory platform for learning musical intervals that contains auditory, haptic, and visual representations.
    \item An open-sourced learning interface that orchestrates the different sensory modalities.
    \item An evaluation of the haptic device through a spatial discrimination study.
    \item An initial study with 18 participants that shows a spatial mapping for haptics can be learned and integrated without additional training. 
    \item Pre-training and post-training tests that allow us to evaluate learning using our device and web platform.
    \item Design recommendations for multisensory systems for perceptual learning tasks.
\end{itemize}

\section{Background \& Related Work}

Our perception relies on the ability to interpret various stimuli through our senses. In particular, auditory perception plays an important role in our daily lives, allowing us to distinguish between sounds and understand their meaning. Such auditory skills can often be refined through focused training.

In this work, we focus on musical ear training as an example of learning an auditory perceptual skill. We first discuss the background of musical ear training, with a focus on musical interval recognition, and then we discuss learning methods and end with ones with haptic feedback.

\subsection{Musical Ear Training \& Pitch Interval Recognition}

% [talk about musical ear training in general. for example jazz musicians. there is also perfect (absolute) pitch recognition, etc etc ]
Ear training, or aural skills training, involves learning to identify musical elements like melodies, chords, rhythms, and their building blocks by how they sound. This centuries-old practice remains a core aspect of Western musical pedagogy, extending beyond early education to even the college level~\cite{karpinski2000aural}. The ability to rapidly recognize musical structures is essential for many musicians, for example jazz musicians rely on this ability to interpret and respond to others in improvisational contexts.

Ear training often begins with learning to identify the distances between notes in a melody as this is considered ``basic to good musicianship''~\cite{buttram1969influence}, and the distance is known as a musical interval. We have selected pitch interval recognition for a number of reasons we discuss below.

Pitch interval identification is a well-studied perceptual task~\cite{siegel1977absolute,zatorre1979identification,miyazaki1992perception}. Many studies have been conducted on pitch interval perception and it has been established as a learnable but also challenging perceptual skill~\cite{little2019inducing}. 
Trained musicians perceive musical intervals as discrete categories. In contrast, many novices perceive musical intervals as being perceptually ambiguous \cite{burns1978categorical}.

Unlike a perceptual task fabricated for a scientific study, pitch intervals are a real skill. As noted, ear training is an integral part of musical training and supports both fundamental pedagogy~\cite{pesek2020troubadour} and advanced activities like transcription~\cite{fletcher2019virtual}. Pitch interval identification is a skill that is difficult to pick up without training. Although most humans have the ability to recognize precise frequencies \cite{jhm2010} pitch interval recognition is still a challenging task. This is useful because it ensures that participants are starting from a similar skill level.  Pitch interval recognition is in a class of tasks that are unpleasant to learn. Students often dislike learning musical intervals because the traditional rote approach to learning is repetitive and not engaging. Thus, techniques for making such learning more enjoyable or easier might indirectly improve learning outcome. 

\begin{figure}[b!]
    \centering
    \includegraphics[width=\linewidth]{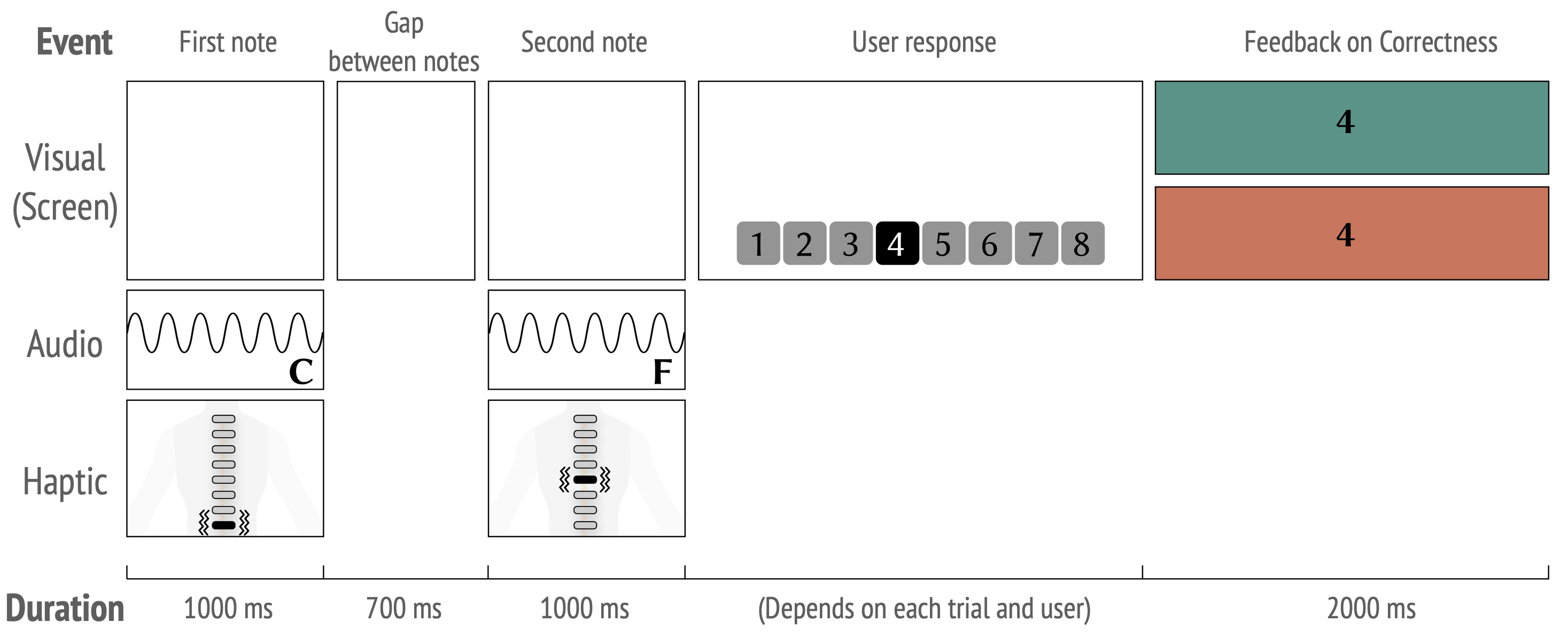}
    \caption{The user experience procedure of a single trial. The user hears (and feel) two musical tones in sequence with a small gap in between. Then they type a number representing their guessed interval and receive on-screen feedback about their guess.}
    \Description{xxxx}
    \label{fig:procedure}
    \vspace{-8px}
\end{figure}

\subsection{Haptic Interfaces for Auditory Perceptual Learning}
% \subsection{Multisensory Learning \& Haptic Interfaces}
\label{percept}

Perceptual learning requires sustained attentional focus and noticing minute details. Multisensory interfaces and contexts have a high potential to improve learning. A meta-analysis discusses three possible mechanisms by which a multisensory context may improve perceptual learning: 1. improved perceptual decision-making; 2. increased saliency and attentional focus; 3. additional or more reliable information \cite{li2022facilitation}. These mechanisms highlight why multisensory interfaces are a particularly good fit for perceptual learning.

% [give some examples of prior work] 
Much of previous work on perceptual learning has focused on audio-visual learning \cite{li2022facilitation}. However, multisensory learning with haptics presents an exciting opportunity because the tactile sense can be highly salient and memorable \cite{hutmacher2018long}. Haptics have been used effectively to aid implicit motor learning. Passive haptic learning (PHL) is a technique that helps individuals learn a motor task without the repetitive experience of rote learning. Prior work has used this technique to teach individuals piano \cite{seim2015towards}, Morse code \cite{seim2016tactile}, and braille \cite{seim2014passive}. Follow-up studies have discussed how PHL is a misnomer as the learning is not passive but active and implicit \cite{pescara2019reevaluating, donchev2021investigating}. Our work aims to achieve a similar effect of implicit learning where users do not experience the repetition of memorizing. Additionally, previous work has found that a visual-haptic device was able to aid in learning of an arm movement more accurately; however this information was not retained after a 4 day rest period \cite{bark2014effects}.

More related to our work is the use of haptics to guide pitch correction. \textit{HapTune} relies on the spatiotemporal placement of two vibrotactile actuators to provide pitch errors to novice violin players. The authors showed using an intuitive metaphor of height required virtually no training for perception and recall\cite{yoo2014initial}. Their longitudinal study result also suggests using haptic feedback caused less visual distraction during instrument play \cite{yoo2017longitudinal}. Similar to HapTune, we leverage the spatial placement of haptic feedback to convey the notion of a big or small musical interval.  Next, we describe the design and implementation of our multisensory learning platform.

Our paper addresses a similar topic of musical intervals, but focuses on a different problem. Yoo and Choi (2017) provide a haptic alternative to visual tuners and evaluates its potential. Our work focuses on using a multisensory interface to improve perceptual learning efficacy.

\section{Design of Multisensory Interval Training System: Purrfect Pitch}

We set out to design a multisensory system that allows novice learners to identify musical intervals beyond rote training. 
Specifically, we have the following design goals: First, we want to make sure the system can improve the learning outcome in terms of the accuracy of the users' response.
Second, usability. The system usage should be intuitive, and the additional haptic stimuli should not distract the learner. The wearable device should be comfortable to wear over time. 
Third, the system should be engaging, and novice learners should prefer to train with it over rote practice, as motivation plays an especially significant role in perceptual learning.
Finally, we want to enable the community to build on top of our tools, and thus we have open-sourced our designs for future work to build upon.

We designed a multisensory platform consisting of a wearable device that provides haptic feedback, and a digital interface that plays musical notes and displays information about the learner's performance. We chose the Western diatonic (major) scale and only evaluated ascending intervals, as it is a common starting point in Western ear training lessons. This gave us eight possible intervals. Additionally, we limited possible tones to mid-range frequencies between C2 (65.4 Hz) and B4 (493.88 Hz), and the training set combinatorially looped within this range. 

An example of a user’s experience is shown in Figure \ref{fig:procedure} and in the Video Figure. The training experience is as follows: Two tones of a randomly selected interval (between 1-8) are played sequentially for each trial. We chose a note length and vibration length of 500ms based on the average length of vibration in \cite{seim2015towards}. At the same time, the user feels vibrations sequentially in two places: first always from the bottom-most module near the lower back, then from another module certain distance (or "interval") apart. For example, if the trial was an interval of 8 (an octave), the module towards the lower back vibrates when the first note is played, and when the second note is played, the wearable system vibrates the eighth module from the lowest one (i.e., the top most module near the neck). After hearing and feeling the notes, the user responds by pressing a number key from 1 to 8. The interface then displays the correct interval number and a green or red screen for a correct or an incorrect response, respectively. The subsequent trial would again choose a random interval and use the second tone of the previous trial as the new first tone. Next, we describe the design considerations of the different components of our platform.

\subsection{Hardware Design}

Prior work shows the natural mapping "between the body position and egocentric orientation" is an intuitive way to guide an individual's attention \cite{choi2012vibrotactile}. We chose a vertical arrangement of the haptic modules to match the vertical spatial metaphor used in Western music, where higher pitches are perceived as "going up" or placed higher on the body. Following this spatial metaphor, we used the eight modules to map to the eight intervals (i.e., eight possible spatial differences) in an octave. We used the location of modules to encode interval information and kept the vibration pattern or intensity consistent. We considered several body study areas: back, forearm, and wrist. Ultimately, we chose the back because it has a large surface area, and thus the spatial difference between neighboring modules is more pronounced.

\begin{figure}[b!]
    \centering
    \vspace{-15px}
    \includegraphics[width=\linewidth]{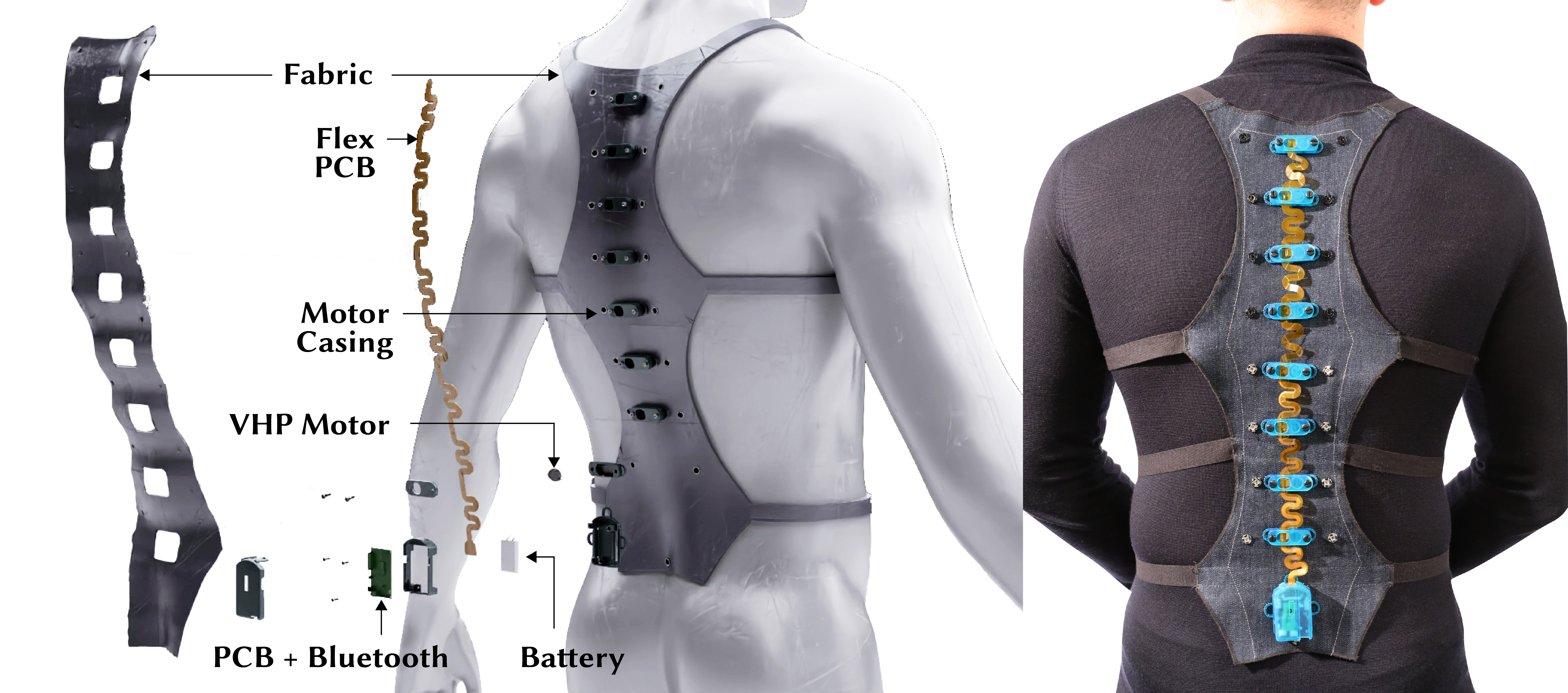}
    \caption{Overview of the haptic device. It is a vest like wearable with eight vibrotactile modules evenly distributed along the spine. The haptic device can be controlled by our web-based learning interface via Bluetooth.}
    \Description{xxxx}
    \label{fig:hardware}
\end{figure}

We built on the vibrotactile haptic platform (VHP) used for on-body haptic research \cite{dementyev2021vhp}. We redesigned the flexible PCB layout to conform to the back along the spine. We modified the physical enclosure of the modules so that they can be stitched onto a piece of fabric (Figure \ref{fig:hardware}). The Bluetooth module of the VHP wirelessly communicates with our learning interface running on a PC. The distance between two neighboring modules is 3cm, and the total length of the electronics is 75 cm.  
We followed the findings by Plaisier et al. \cite{plaisier2020perception} on individuals' perceptual distance of vibrotactile simulations around the spine. In their findings, the vertical layout resulted in the least overall variance of perceptual distance, and we chose the 3cm to fit all eight modules on the spine. Since our setup does not exactly replicate Plaisier et al.'s work, we also conducted a perceptual study to understand the spatial discrimination between the modules, detailed in Section \ref{sec:spatial}.

% To optimize for spatial discriminability, our placement on the back was able to pass the two-point threshold \cite{choi2012vibrotactile}. 

\subsection{Learning Interface}

As discussed in section \ref{percept}, perceptual learning cannot be performed passively – it requires active learning with attention to detail and repetition. We used a similar method of implicit active learning and sensory integration utilized in the work by Seim et al. \cite{seim2015towards}. 

The goal of the learning interface is two-fold: (1) to orchestrate the audio and haptic feedback stimuli in different training and test conditions, and (2) to provide the user feedback on the correctness of their response to enable active learning.

The audio and haptic stimuli are played concurrently to enable sensory integration while the screen displayed nothing to limit visual interference effects \cite{li2022facilitation}. The system plays the lower note first followed by the higher note. After the user enters a number, they are given feedback about whether they gave the right or wrong answer, and they see the correct interval number (Figure \ref{fig:interface}). We choose to display a uniform color block of red or green instead of words (“correct” or “incorrect”) to allow users to advance quickly through the trials to avoid requiring extra cognitive processing of the feedback. Users can repeat a "question" by pressing the space key and hear the notes again, but they only have one chance to answer. Two seconds after the user guesses, the next trial automatically advances. 

Setting up the learning system is simple: the web-based interface can be easily deployed and connects to the haptic vest via Bluetooth. The interface offers different toggles for turning on/off the sensory feedback channels.

We have make available the interface in our github repo: \url{https://github.com/<Anonymized for review>}

\begin{figure}[b!]
    \centering
    \includegraphics[width=.75\linewidth]{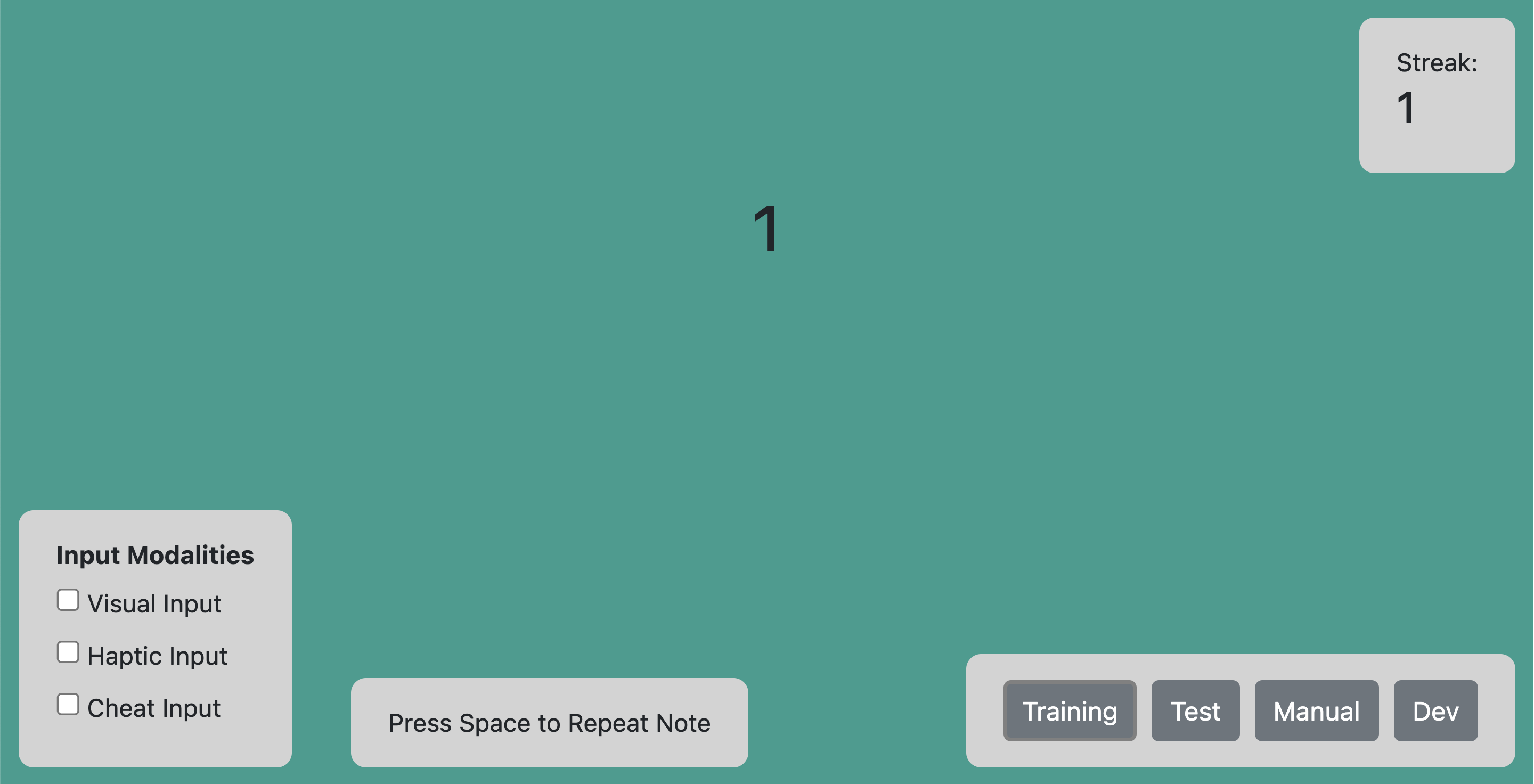}
    \caption{The web-based learning interface that plays the musical notes and connects to the haptic device. The color green represents a correct answer, with a number displayed representing the correct interval number.}
    \Description{xxxx}
    \label{fig:interface}
\end{figure}

\begin{figure*}[htbp]
    \centering
    \includegraphics[width=1\textwidth]{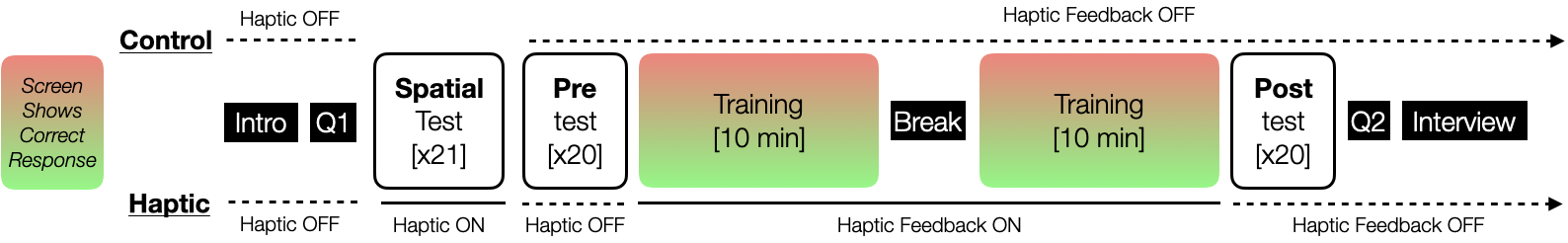}
    \caption{Experimental protocol. The top lines represent the control condition and the bottom lines represent the experimental condition. The dashed lines represent when haptic feedback is absent and a solid line represents the presence of haptic feedback. The numbers in brackets represent the number of trials.}
    \Description{xxxx}
    \label{fig:protocol}
\end{figure*}

\section{Methods}

We conducted a between-subjects in-lab study to evaluate and understand our system as (1) a multi-modal feedback system and (2) a learning tool. Specifically, our research questions are as follows:
\begin{itemize}
    \item RQ1: What is the \textbf{perceptibility} of the spatial distance between two stimuli if participants are only given haptic feedback or only given audio feedback?
    \item RQ2: Does haptic feedback improve the \textbf{accuracy} and \textbf{response time} of identifying musical intervals \textit{during training}?
    \item RQ3: Does the introduction of haptic feedback affect people's cognitive load \textit{during training}?
    \item RQ4: Finally, after training with our system, can novice learners \textbf{learn} musical intervals identification by ear only?

\end{itemize}

\subsection{Experimental Protocol}

\subsubsection{Conditions}
Our experiment consists of the control (without haptics) condition and the experimental condition (with haptics). In the haptic condition, those participants wore the wearable on their back and felt haptic feedback while hearing the musical notes during the spatial perceptual experiment and the training sessions. We conducted a between-subjects experiment to avoid learning effects from one condition to another within a participant. 

\subsubsection{Participants}
We recruited 18 participants (age 18-36; 5 identifying as female or non-binary) through email and Slack promotion. Our participants consisted of undergraduate and graduate students who reported having normal hearing, no to little prior training in identifying musical intervals, and none to minimal experience with playing musical instruments. Participants were recruited through our university mailing lists. The participants were randomly assigned to either the audio-only control (n = 10) or the audio-haptic condition (n = 8). The study lasted around 45 minutes, and the participants were compensated with \$15 in the form of a gift card upon completion.

\subsubsection{Procedure}

Figure \ref{fig:protocol} shows the experimental procedure. All participants were asked to sit in front of a computer and given over-the-ear headphones (Sennheiser HD 560) to wear for the duration of the experiment. Only participants in the haptic condition were given our wearable device to wear on the back over tight clothing.

We start the experiment by first explaining the concept of musical intervals. The participants then filled out a demographics questionnaire (Q1).  The participants in the haptic condition also went through an additional spatial perceptual experiment. The spatial perceptual experiment aims to measure how well a person can identify the felt spatial distance between two sequential vibrations on the back along the spine. There are eight possible pairs of vibrations, all starting with the bottom-most module closest to the tailbone. The user will respond to each pair of vibrations by typing a number that represents the felt spatial distance between the vibrations. They could use a non-zero, positive number, subject to the constraint that a larger number should correspond to a greater distance. The order of the eight pairs of vibrations is randomized. The experiment ends when all eight pairs of vibrations have been played.

For both conditions, participants used our learning interface in which they must rapidly identify pitch intervals. They were asked to go through as many trials as possible. The user response, number of correct responses, response time, and number of repeats were collected. To avoid fatigue, participants went through two ten-minute training sessions with a small break in between. Only participants in the haptic condition felt haptic feedback during the training sessions.  We conducted a short test before and after the training sessions (a pre-test and a post-test). The tests aim to measure the participants' true auditory perceptual ability, meaning that they were done without haptics and without on-screen feedback on the learning interface about the correctness of their response.

Finally, at the end of the study, participants completed a brief questionnaire (Q2) containing NASA TLX (Task Load Index) questions and additional questions about such as levels of engagement and effectiveness. They also went through a semi-structured interview about the learning experience and shared any other comments with the experimenter.

\begin{figure}[b!]
    \centering
    \includegraphics[width=1\linewidth]{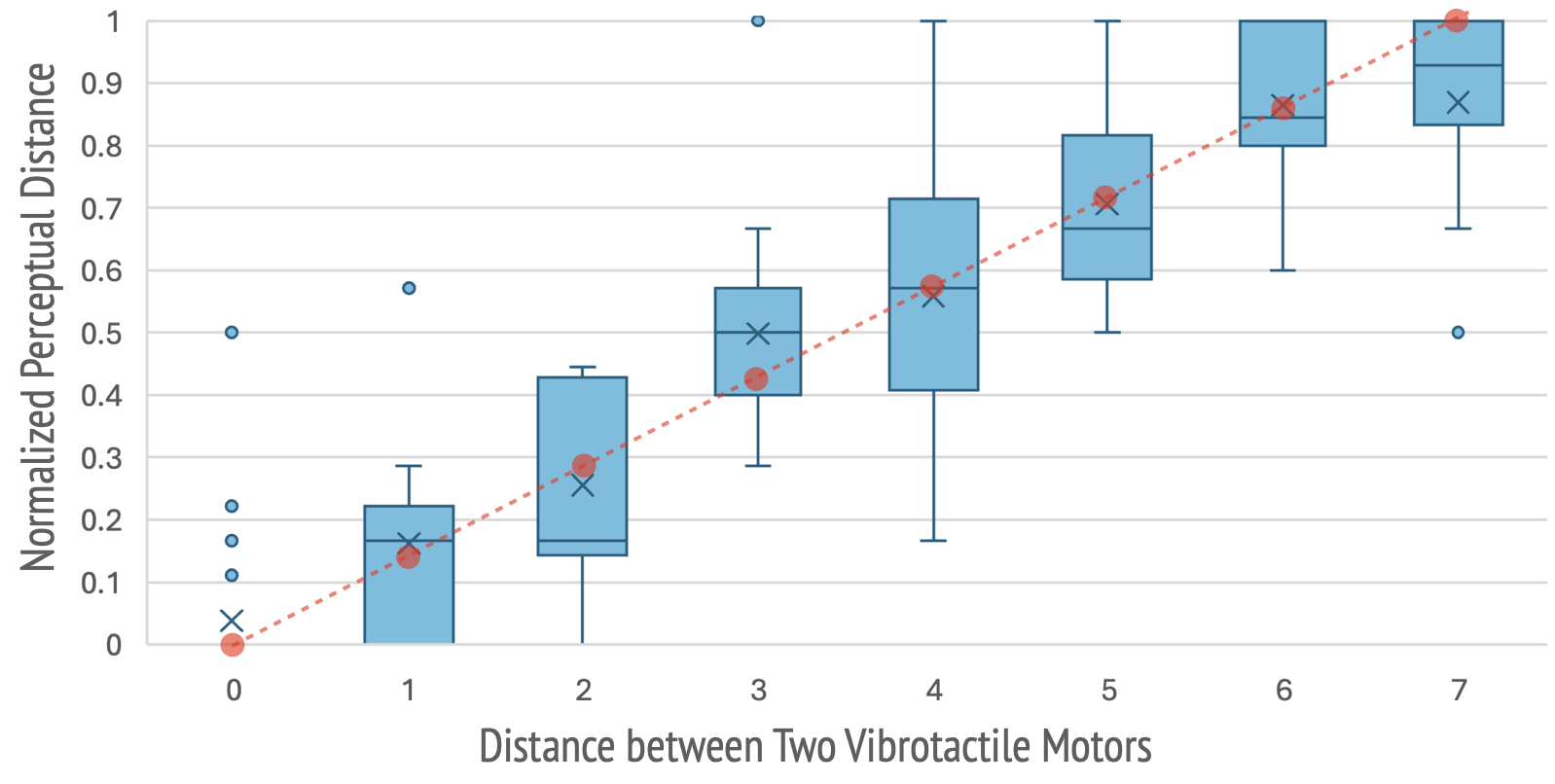}
    \caption{Box plots of the perceived distance between two vibrotactile motors on the back. The cross mark indicates the mean value. Red dots and the red dashed line indicate the ground truth.}
    \Description{xxxx}
    \label{fig:spatial}
\end{figure}

\section{Results \& Analysis}

Here we represent results and analysis from the quantitative experimental and survey results and qualitative responses to answer each of our research questions.

\subsection{Perceptual Estimation of Spatial Distance}\label{sec:spatial}

First, we want to understand the perceptibility of the spatial distances of each pair of haptic modules on the back. This part of the study is similar to the free magnitude estimation test \cite{snyder2021assessment, plaisier2020perception}, but we are interested in the relative distance within a distance range determined by our hardware rather than the absolute scale.

Similar to the setup of a free magnitude estimation task, we asked participants to rate the felt distance on a scale of their choice. To account for the variations in individuals' range of distance values, we first normalize each participant's score by fitting their ratings to a 0-1 scale, where 0 is the minimum rated distance, and 1 is the maximum rated distance. Concretely, within a participant, for each score, we divide the difference between the score and the minimum value over the difference between the maximum and minimum values. 

We first want to know if the perceptual distances of each pair of motors are distinct from each other, especially the neighboring pairs. Figure \ref{fig:spatial} shows the participants' distance estimation of the 8 pairs of vibrations. As expected, the perceived distance increases as the actual distance between the two haptic motors increases.
Then we look at whether the neighboring distances can be differentiated from each other. The differences between the neighboring distances are  mostly clear, except for between distances 3 and 4 and  between distances 6 and 7. We then look at the direction in which the estimated distances deviate from the ground truth. The red dots and the red dashed line indicate the ground truth distance responses. Specifically, distance 3 is estimated to be bigger and distance 7 is estimated to be smaller. The top motor's position for distance 7 is near the neck, farthest away from the bottom motor. 
It is also worth noting the significant variance of distance 4. This shows that distances where the top motor is around the middle of the spine are notably perceptually ambiguous. 

% The uncertainties in the haptic modality may influence the perceived auditory distance during training.

% [TODO, add graph with comparison chart-  In short, the results suggest that the haptic stimuli helped participants to be significantly better at identifying the correct internals. And even if they guessed incorrectly, the guessed responses were closer to the correct response than those in the control group.]

\begin{figure}[t!]
    \centering
    \includegraphics[width=.75\linewidth]{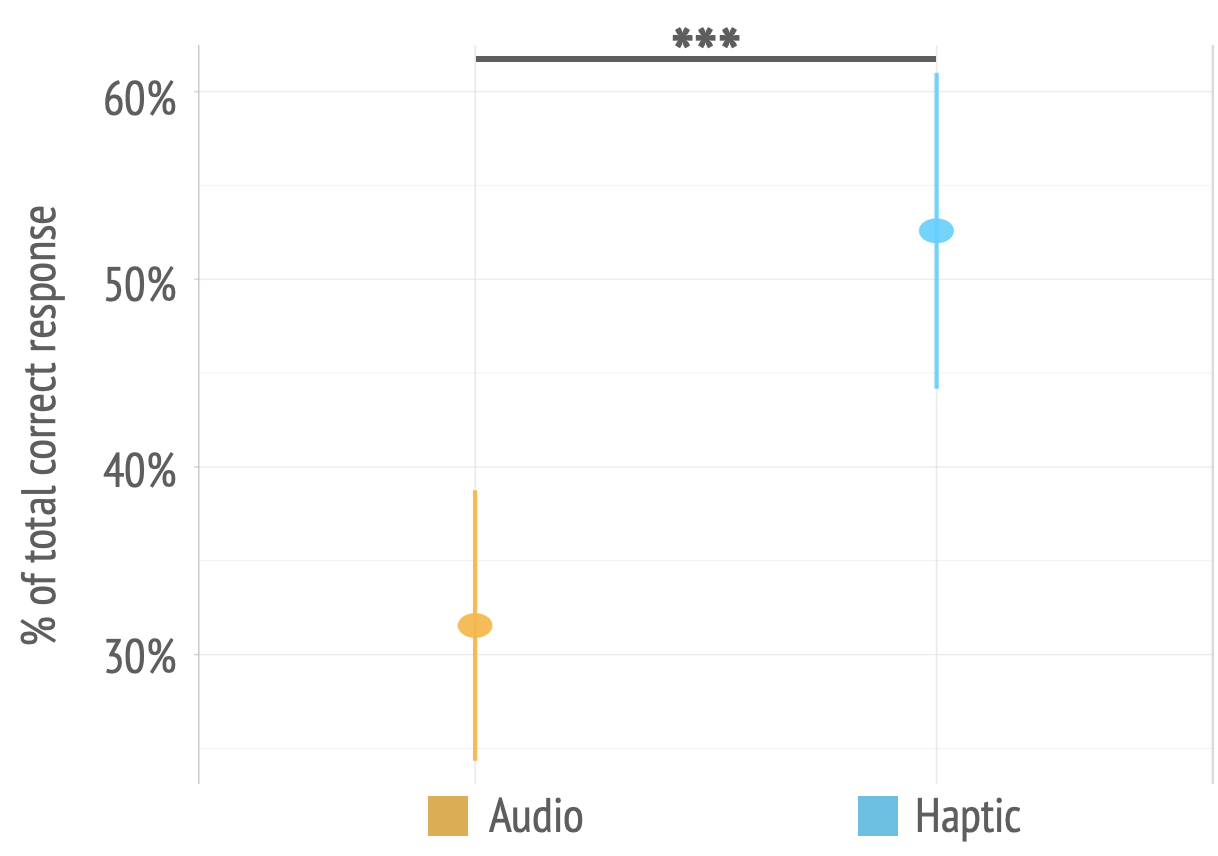}
    \caption{Total percent of correct response per condition during training based on mixed-effects logistic regression. The adjusted marginal prediction of a correct response is 34\% (95\% C.I.=[0.270, 0.410]) for the audio-only group and 54.3\% (95\% C.I.=[0.463, 0.624]) for the audio-haptic group. *** p<0.001}
    % \caption{Correct based on mixed-effects logistic regression \normalfont 
    % The adjusted marginal predictions of a correct response was 34\% (95\% C.I.= [0.270,  0.410]) for the audio-only group was and 54.3\% (95\% C.I.= [0.463, 0.624]) for the audio-haptic group.}
    \Description{xxx}
    \vspace{-15px}
    \label{fig:expected-values}
\end{figure}

\subsection{Accuracy, Learning Rate and Response Time during Training Phase}

\textbf{\textit{Effect of haptic stimuli on accuracy during training}}

In Research Question 2, we want to understand if users would perform better when they receive both auditory and haptic feedback. To evaluate this, we performed a mixed-effect logistic regression (binomial family generalized linear mixed-effects model) in R using the \textit{lme4} package \cite{bates.etal_2015}. Such models are a common analysis method in psychophysics for perceptual tasks \cite{moscatelli.etal_2012}. We based our regression on the data collected from participants during the "Training" phase which resulted in 3052 observations total across groups. We modeled the likelihood of getting a question correct based on haptic input (a binary variable to indicate whether haptic input was present or not), the number of trials completed (to account for improvement over time), and the interaction between haptic input and the number of trials (to examine whether the rate of improvement/decline differed between the haptic and audio-only conditions). We accounted for individual differences with a random effect for participant ID. 

We calculated the marginal adjusted predictions of a correct response for the audio-only group as 34\% (p<0.001, 95\% C.I.=[0.270,  0.410]) and audio-haptic group as 54.3\% (p<0.001, 95\% C.I.=[0.463, 0.624]) using \textit{marginaleffects} \cite{arel-bundock_2024}. The difference in predictions can be seen in Figure \ref{fig:fit-correct}. Adding haptic input to auditory input increases the likelihood of correctness by about 20\% on average. We performed a hypothesis test on this contrast and found this accuracy difference was statistically significant ($0.203$, p<0.001, 95\% C.I. = [0.0967, 0.31]).
\begin{figure}[t!]
    \centering
    \includegraphics[width=0.75\linewidth]{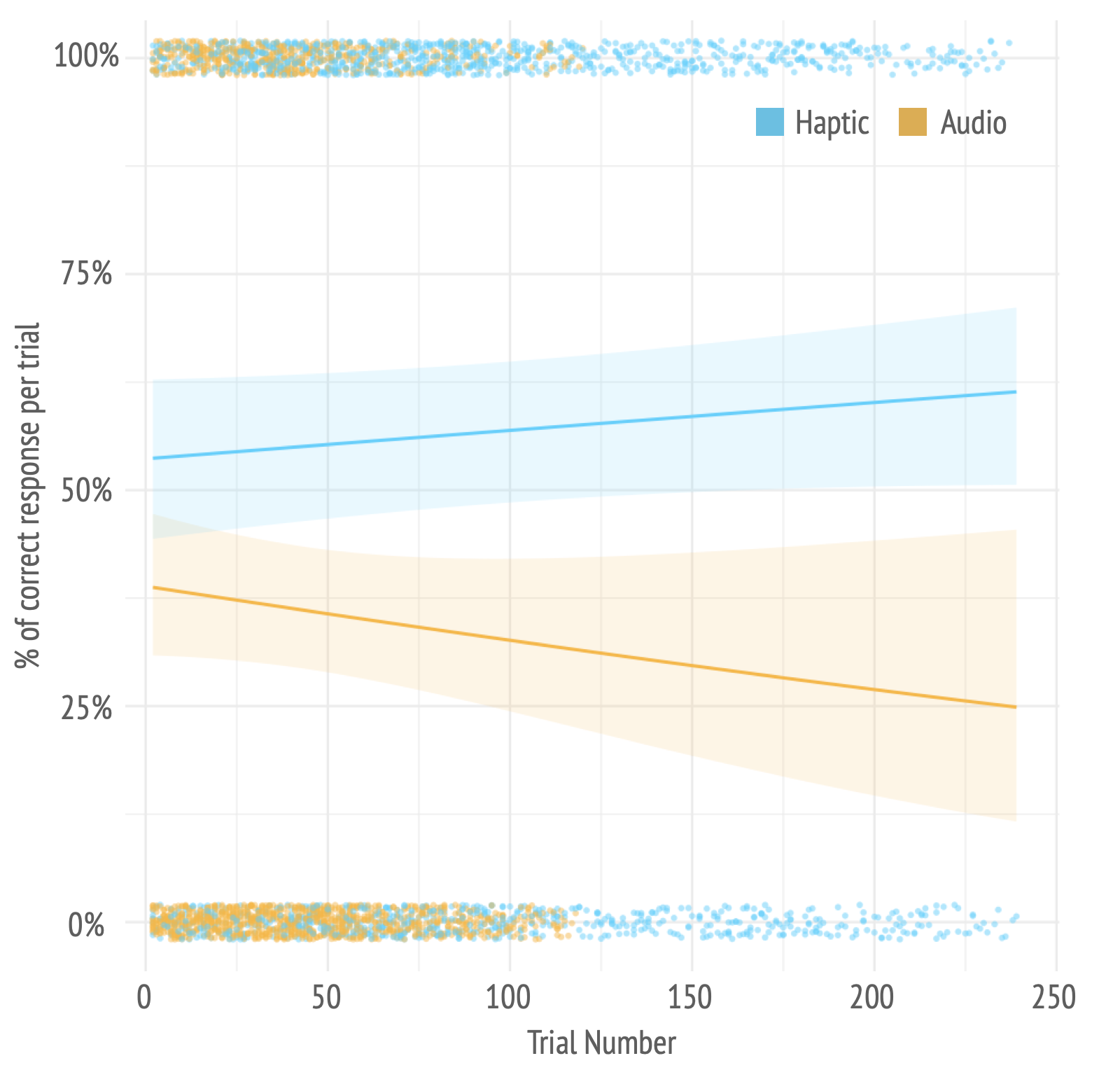}
    \caption{Percent correct response per trial over the number of trials during training. The dots represent the raw data. The lines and confidence intervals are average marginal predictions of the mixed-effect logistic regression.}
    \label{fig:fit-correct}
    \Description{xxxx}
\end{figure}

\textbf{\textit{Effect of haptic stimuli on learning rate}}
Additionally, we want to compare the learning rate for both groups. We define the learning rate as how much the user is expected to improve for one additional trial (change in expected likelihood of correctness over the change in trial number). Figure \label{fig:fit-correct} shows a slightly positive learning rate (3.17e-4, p =0.172, 95\% C.I. =[-0.000138, 0.000771]) for the haptic-audio participants and slightly negative (-5.92e-4, p=0.218, 95\% C.I.=[-15.34e-4, 3.50e-3]) for the audio-only participants; however, both of these results are statistically insignificant. The average learning rate is slightly negative (-1.19e-4, p=0.647, 95\% C.I.=[-6.31e-4, 3.92e-4] which is unexpected.

\begin{figure}[t!]
    \centering
    \includegraphics[width=.75\linewidth]{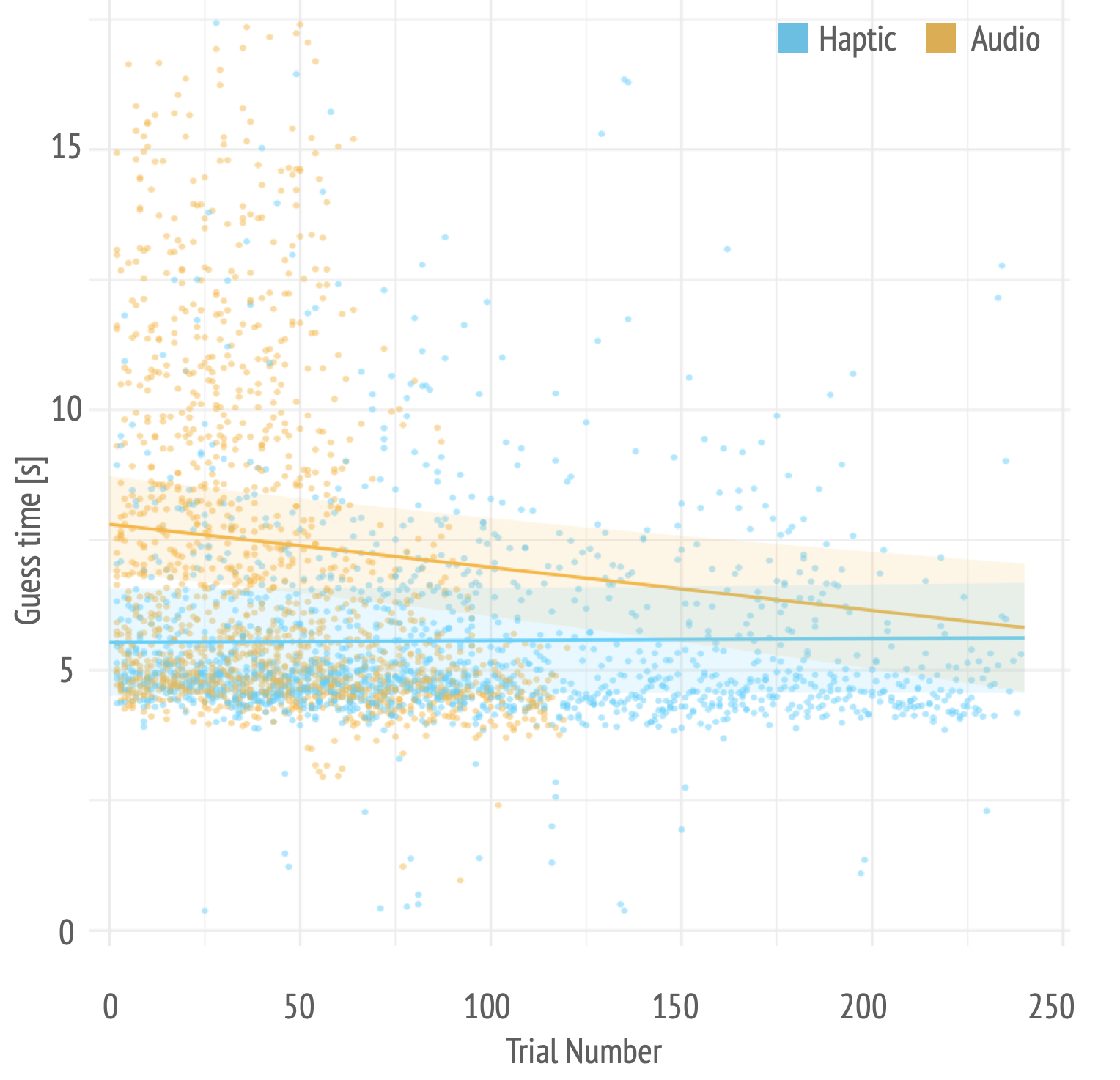}
    \caption{Guess time per trial over the number of trials during training. The dots represent the raw data. The lines and confidence intervals are average marginal predictions of the mixed-effect linear regression.}        
    \label{fig:guess-time}
    \Description{xxxx}
\end{figure}

\textbf{\textit{Effect of haptic stimuli on response time}} 
In addition to response accuracy, we also investigated the effect of introducing haptic feedback on users' response time (i.e., the time difference between the onset of the first note in the interval and when the participant pressed a number key).  We dropped times below 1.2 s, which is the earliest a user could hear the second note because anything before that would be effectively a random guess or accidental entry. We also dropped outliers beyond 2$\sigma$ (17.2s) – most of these long guess times involved the user asking a question and were not considered a valid trial.  Figure \ref{fig:guess-time} shows the mean response time over the course of the study for each condition. We performed another mixed-effect linear regression, modeling the response time based on haptic input, the number of trials, the interaction term, and a random effect for participant ID. 

The audio-only group has a higher predicted average response time (6.918s, p<0.001, 95\% C.I.=[6.001, 7.836]) compared to the haptic-audio group (5.244s, p<0.001, 95\% C.I.=[4.229, 6.259]).  We performed a post-hoc contrast to compare these (audio-only vs. haptic-audio), and found a statistically significant difference of 1.674s (p=0.0165, 95\% C.I. =[0.306, 3.043]). We also see that the slope for the audio-only group is negative (-0.00828, p < 0.001, 95\% C.I.=[-0.0012.4,  -0.00415]) which indicates that audio-only participants got faster over time. The haptic-audio group has a slope that is comparatively flat (0.356, p =0.72086, 95\% C.I. = [-0.0016, 0.00231]).

\subsection{Comparing accuracy during pre-test and post-test}
\begin{figure}[t!]
    \centering
    \includegraphics[width=.75\linewidth]{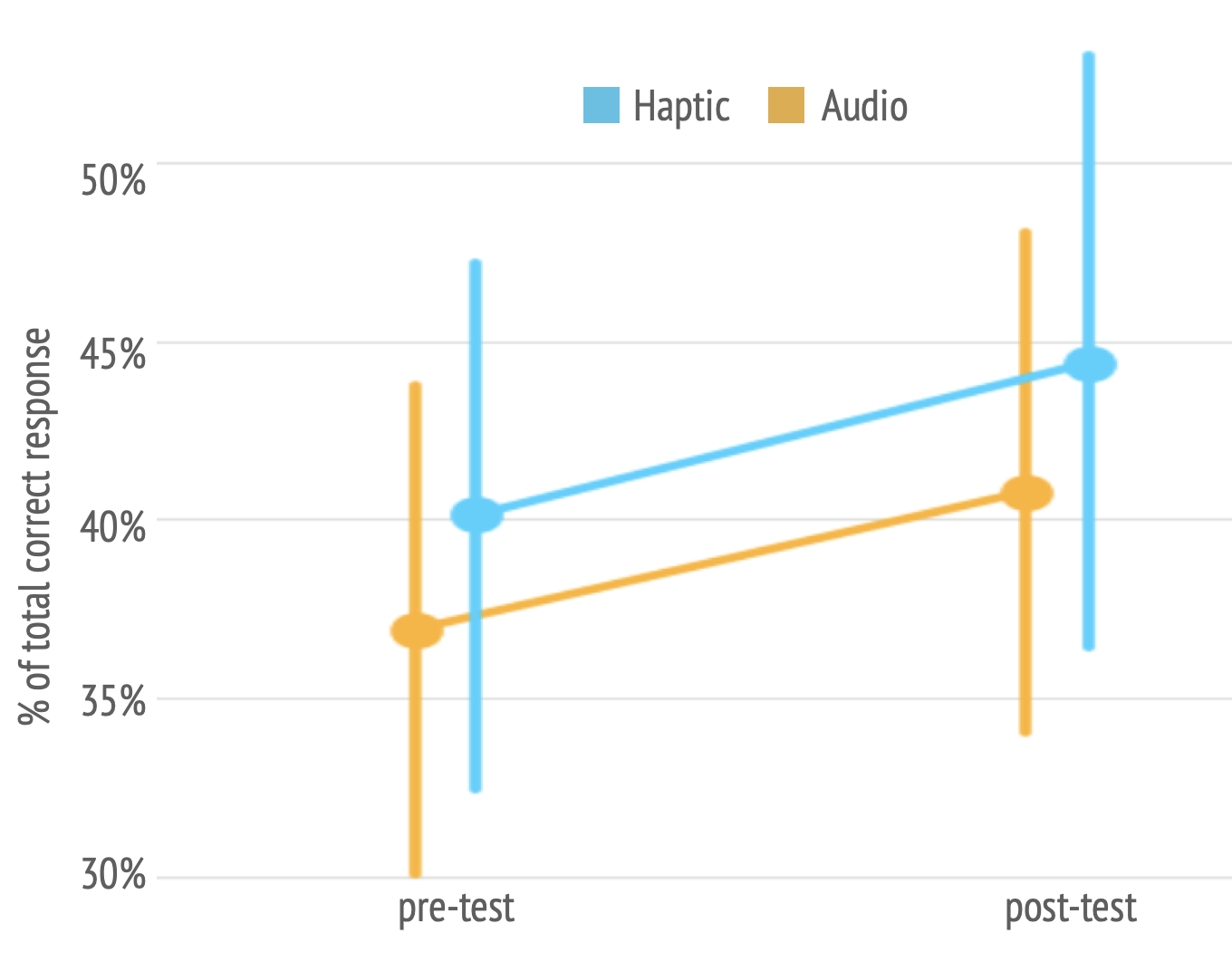}
    \caption{Accuracy of pre-test and post-test. Users who received audio-only training got on average 37\% correct on the pre-test (95\% C.I.=[0.297, 0.441]) and 41\% on the post-test (95\% C.I.=[0.335, 0.481]). Users who received haptic-audio training got on average 40\% correct on the pre-test (95\% C.I.=[0.324, 0.479]) and 44\% on the post-test (95\% C.I.=[0.364, 0.524]).}
    \label{fig:pre-postv2} 
    \Description{xxxx}
\end{figure}

We also measured learning improvement by comparing the scores from the pre-test and post-test. For both tests, each group received 20 audio-only questions without feedback on whether they responded correctly. Users who received audio-only training got on average 37\% correct on the pre-test ($\mu$=0.369, $\sigma$=0.036, 95\% C.I.=[0.297, 0.441]) and 41\% on the post-test ($\mu$=0.408, $\sigma$=0.037, 95\% C.I.=[0.335, 0.481]). Users who received haptic-audio training got on average 40\% correct on the pre-test ($\mu$=0.401, $\sigma$=0.039, 95\% C.I.=[0.324, 0.479]) and 44\% on the post-test ($\mu$=0.444, $\sigma$=0.041, 95\% C.I.=[0.364, 0.524]). In Figure \ref{fig:pre-postv2}, the change in percentage correct is about the same for both groups($\Delta_{audio} = 0.0385, \Delta_{haptic} = 0.0424$), and the difference is not statistically significant (t=0.03, p=0.9747).

\subsection{Cognitive load during training}

We administered a partial NASA TLX questionnaire \cite{hart1988development} and a separate set of questions. The full questionnaire and the original rating scale can be found in the Appendix. Figure \ref{fig:tlx} shows the questionnaire results. For readability, we converted the scores for this figure, where a higher score is more desirable (e.g., a higher score means the task is \textit{less} mentally loaded and hence more preferred).  We found that across the board, the haptic condition performed better. 
% This is expected as discrete haptic vibrations on the body are more salient to perceptually identify than audio frequencies.  

The frustration score for the haptic condition is significantly lower (paired t-test, p<0.05). It is worth pointing out that learning a new skill presents challenges beyond improving one's performance, such as maintaining one's motivation \cite{turakhia2021adapt}; in our case, learning musical intervals typically requires arduous rote ear-training. Figure \ref{fig:tlx} shows that participants reported feeling more engaged, and considered the learning experience with haptics more effective and fun. Anecdotally, participants in the haptic condition reported that they “relied more on instincts during haptic trials” (P8) and were “more confident when pressing the key (to give their response)” (P3).

\begin{figure*}[hbtp]
    \centering
    \includegraphics[width=.75\linewidth]{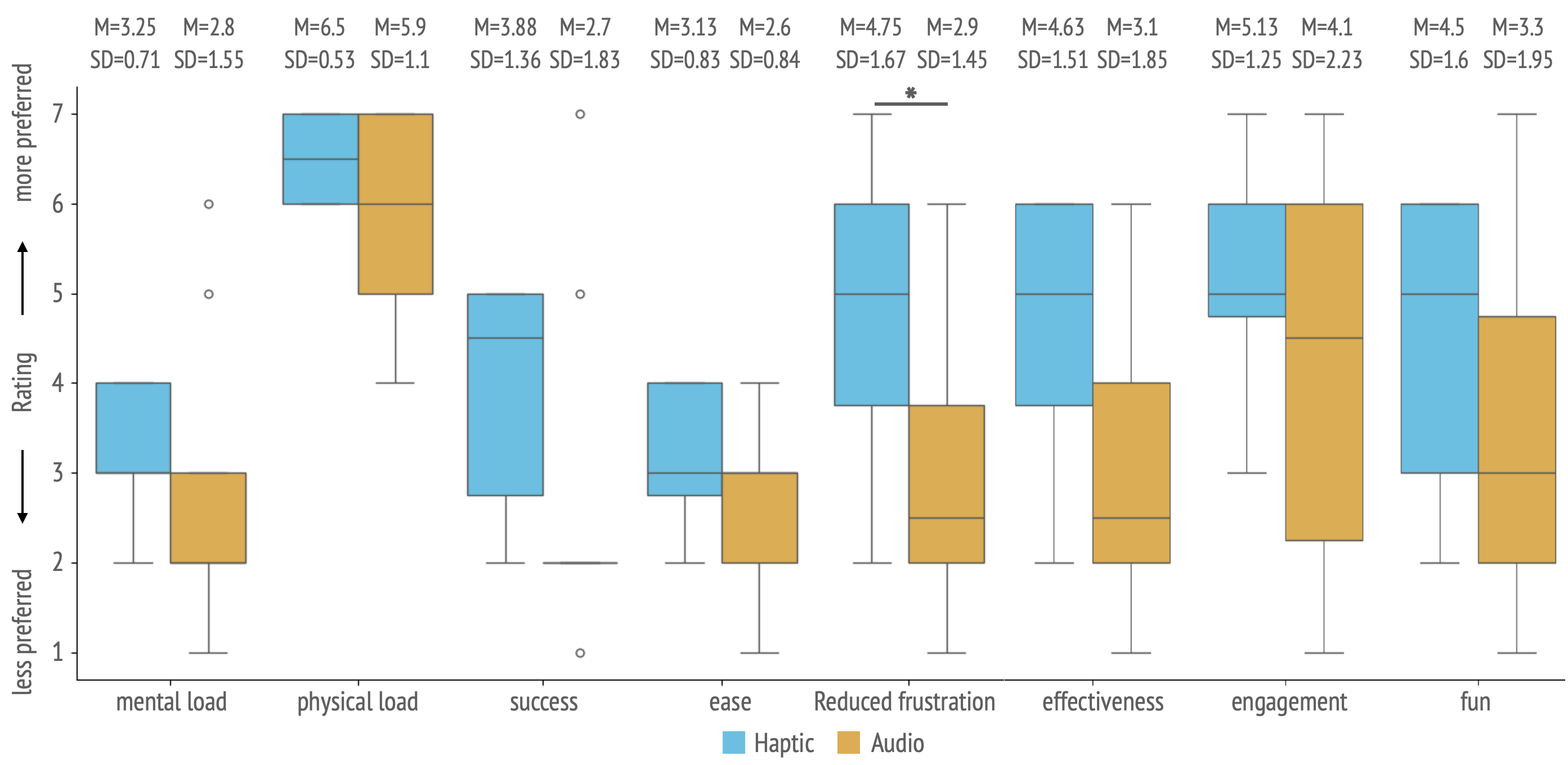}
    \caption{Box plots of the questionnaire results. M: mean, SD: standard deviation. *: p<0.05}
    \label{fig:tlx}
    \Description{xxxx}
\end{figure*}

\section{Discussion}

% \subsubsection{Our multisensory system clearly improves accuracy and response time for the musical interval identification task.}

Overall, we found that participants in the auditory-haptic experimental condition performed better compared to the audio-only group during training. Our primary findings are: \textbf{(1) The haptic-auditory condition performed 20\% more accurately compared to the audio-only control.} \textbf{(2) The haptic-auditory condition responded 1.674s faster than the audio-only control.} 

We suspect that one reason haptic participants were able to perform much better during training is purely on an information-theoretic basis – they simply had more information to inform their guess. For example, Participant 2 reported that haptic feedback “allowed me to narrow down my interval guess within a range”. Additionally, participants in the haptic condition commented on noticing how their sensory weights changed as their auditory perception improved: “[I] use the haptic distance first and connect it with audio” (P1) and “Coming into this, [I was] relying exclusively on haptics. [I] started to notice patterns in sound.” (P7). The spatial discrimination test (Figure \ref{fig:spatial}) also shows that the design and placement of our haptic device on the back produced sufficiently distinct stimuli that benefited the judgment of audio intervals.

Both of these results support the idea that multisensory interfaces with haptics have the potential for augmenting perception in a short time (less than an hour). We believe we were able to see such rapid improvement because multisensory integration is a dynamic process which adjusts quickly and subconsciously. Multisensory integration is the mechanism through which the nervous system uses sensory information of varying reliability to create a coherent perception of the world. For example, speech perception is not purely an auditory task - depending on the noise in the environment humans may rely on vision more than sound in order to understand speech \cite{bruns_2019}. Often this process is considered dynamic Bayesian optimization problem in neuroscience \cite{deneve.pouget_2004} because the reliability of a given sense varies depending on the environment.

Even though the haptic group did not significantly outperform the audio (control group) in the pre-post training test results (Figure \ref{fig:pre-postv2}), we were able to find interesting insights when looking at the partcipants' response by interval. 

First, the results show that removing the multisensory device did not \textit{negatively} impact performance in our study. One concern with using technology to augment perception is overreliance - for instance, blind spot indicators are useful, but this overreliance may cause problems when a driver is in a different car. In our case, overreliance would look like users ignoring the audio information and only using the haptic device for the task. However, we did not see a lower in the accuracy count for the haptic group after training.

\begin{figure*}[hbtp]
    \centering
    \includegraphics[width=.75\linewidth]{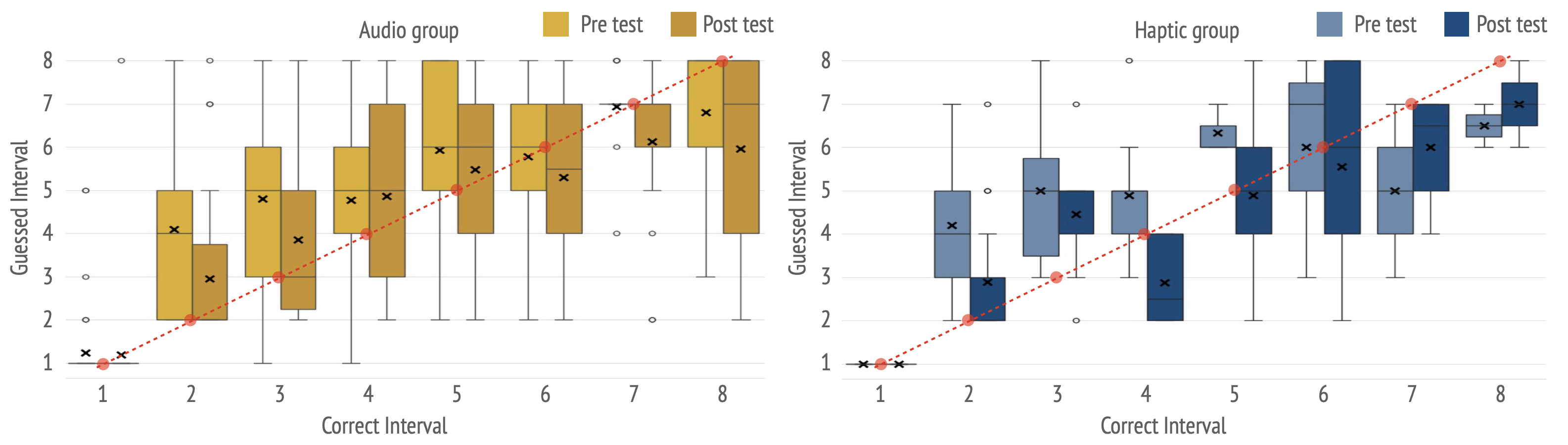}
    \caption{Box plots of the guessed interval for pre- and post- test broken down by interval. The left figure is for the audio control group and the right figure is for the haptic group. The cross mark indicates the mean value. Red dots and the red dashed line indicate the ground truth.}
    \label{fig:spatial_prepost}
    \Description{xxxx}
\end{figure*}

In addition, the haptic device positively influenced participants' judgment of specific intervals that performed worse than others during the pre-test.
Figure \ref{fig:spatial_prepost} shows the pre-post test results broken down by intervals. We see that the haptic group shows lower and reduced variances in post-test guesses across intervals compared to pre-test, whereas the control group shows consistently large variances in post-test guesses and no significant reduction compared to pre-test. The post-test guesses of the haptic group also show less anchoring of pre-test guesses. However, these results are only empirical, and the implications of the observed trends require further analysis.

\section{Limitations \& Future Work}

\textit{Small effect size.} One limitation of this study was the short time period of training. Learning of all kinds inherently takes time, and we were limited by how long we could reasonably ask participants to listen to pairs of musical notes. 

\textit{Confounding variables.} Another challenge was the high variation in the prior acuity of musical notes of the participants, even though we recruited only people who did not have prior proper ear training. In post-hoc exploratory statistics, we found a correlation between experience in music and knowledge of musical intervals with the change in pre- and post-test results. We do not report on these results since they are not a primary contribution to the paper. It is unclear if prior interest and experience in musical tasks reflect a difference in intrinsic motivation or existing knowledge. Future work on perceptual augmentation may wish to explore explicit counterbalancing based on exposure or intrinsic motivation. 

\textit{Personalized learning.} Additionally, the platform could tailor the learning experience to the skill level such as focusing on specific intervals. Some participants mentioned using existing song knowledge as a touchpoint for intervals. Future work could offer tailored haptic feedback to represent specific intervals based on prior knowledge like familiar songs or jingles \cite{levitin1994absolute}. 

\textit{Beyond learning musical intervals.} We believe our wearable device can be used for applications beyond musical interval learning. Future work can investigate leveraging haptics on the back for other tasks involving the metaphor of moving in a vertical dimension, such as learning the tones of tonal languages like Mandarin.

\section{Conclusion}
We designed Purrfect Pitch, a game-based platform for perceptual learning and a wearable for assisting interval identification with haptic feedback. Our initial study with 18 participants revealed augmenting auditory tones with haptic feedback improves accuracy by 20\% and response time by 1.67 s. In addition, participants were more engaged and motivated to learn, which are important aspects of learning. Overall, this line of work suggests the potential of multisensory learning for acquiring a novel perceptual skill (such as musical interval identification). 

% \begin{acks}
% We want to thank our participants for their invaluable time. Additionally thank you to Nikhil Singh and Rose Hegele for providing expertise on interval training pedagogy and musical concepts broadly, and thank you to Abhi Jain and Caitlin Morris for feedback on the paper. The study was approved by the MIT Committee on the Use of Humans as Experimental Subjects (COUHES) 2210000773 under the title: Evaluating the Efficacy of Multimodal and Haptic Feedback on Learning and Recalling Musical Intervals. Barakah as well!
% \end{acks}

\bibliographystyle{ACM-Reference-Format}
\bibliography{sample-base}

\appendix
\section{Questionnaire}
\begin{itemize}
    \item Mental load: How mentally demanding was the task? (1 = Very low, 7 = Very high; results inverted for analysis)
    \item Physical load: How physically demanding was the task? (1 = Very low, 7 = Very high; results inverted for analysis)
    \item Success: How successful were you in accomplishing what you were asked to do? (1 = Failure, 7 = Perfect)
    \item Ease: How hard did you have to work to accomplish your level of performance? (1 = Very low; 7 = Very high)
    \item Frustration: How insecure, discouraged, irritated, stressed, and annoyed were you? (1 = Very low, 7 = Very high; results inverted for analysis)
    \item Effectiveness: This was an effective way to learn. (1 = Strongly disagree, 7 = Strongly agree)
    \item Engagement: This was an engaging experience. (1 = Strongly disagree, 7 = Strongly agree)
    \item Fun: This was a fun experience. (1 = Strongly disagree, 7 = Strongly agree)
\end{itemize}

\end{document}